\documentstyle[bezier]{qcdparis}
\begin{document}
\def \be  {\begin{equation}}
\def \ee  {\end{equation}}
\def \ba  {\begin{eqnarray}}
\def \ea  {\end{eqnarray}}
\def \baa {\begin{eqnarray*}}
\def \eaa {\end{eqnarray*}}
\newcommand \ci [1] {\cite{#1}}
\newcommand \bi [1] {\bibitem{#1}}
\def \lab #1 {\label{#1}}
\newcommand\re[1]{(\ref{#1})}
\def \qqquad {\qquad\quad}
\def \qqqquad {\qquad\qquad}
\newcommand\lr[1]{{\left({#1}\right)}}
\newcommand\lrs[1]{{\left[{#1}\right]}}
\def \Tr {\mbox{Tr\,}}
\def \tr {\mbox{tr}}
\newcommand \vev [1] {\langle{#1}\rangle}
\newcommand \VEV [1] {\left\langle{#1}\right\rangle}
\newcommand \ket [1] {|{#1}\rangle}
\newcommand \bra [1] {\langle {#1}|}
\def \CO {{\cal O}}
\font\cmss=cmss10 \font\cmsss=cmss10 at 7pt
\def\inbar{\,\vrule height1.5ex width.4pt depth0pt}
\def\IC{\relax\hbox{$\inbar\kern-.3em{\rm C}$}}
\def\IZ{\relax\ifmmode\mathchoice
{\hbox{\cmss Z\kern-.4em Z}}{\hbox{\cmss Z\kern-.4em Z}}
{\lower.9pt\hbox{\cmsss Z\kern-.4em Z}}
{\lower1.2pt\hbox{\cmsss Z\kern-.4em Z}}\else{\cmss Z\kern-.4em Z}\fi}
%\defIC{{\hbox{{\rm I}\kern-.5em\hbox{\rm C}}}}
%\def\IZ{{\hbox{{\rm Z}\kern-.4em\hbox{\rm Z}}}}
\def\IR{{\hbox{{\rm I}\kern-.2em\hbox{\rm R}}}}
\def\IP{{\hbox{{\rm I}\kern-.2em\hbox{\rm P}}}}
\newcommand{\as}{\ifmmode\alpha_{\rm s}\else{$\alpha_{\rm s}$}\fi}
\renewcommand{\Re}{\mathop{\rm Re}\nolimits}
\def\Im{\hbox{\rm Im}\,}

\def\numb{\hfill\parbox{35mm}{LPTHE-Orsay-95/76\par
                        hep-ph/9511370  \par
                        November, 1995}}
\pagestyle{plain}
\title{Theoretical Progress in QCD at small $x\ ^\ast$}
\author{G.P. Korchemsky$\,^\dagger$}
\affil{
Laboratoire de Physique Th\'eorique et Hautes \'Energies$\,^+$, \\
Universit\'e de Paris XI, Centre d'Orsay, \\
b\^at 211, 91405 Orsay C\'edex, France
\\[-75mm]
\numb
\\[75mm]
{}
}
\abstract{
I review a recent progress in understanding the Regge asymptotics
in perturbative QCD including the behaviour of the structure
functions of DIS at small Bjorken $x$.
}
\resume{Je pr\'esente un progr\`es r\'ecent dans la compr\'ehension du
comportement asymptotique \`a la Regge de QCD perturbative, incluant
le comportement de la fonction de structure de la diffusion
profond\'ement in\'elastique pour les
petites valeurs de $x_{\rm Bjorken}$.
}
\twocolumn[\maketitle]
\fnm{7}{Plenary talk given at the Workshop on Deep Inelastic Scattering
and QCD, Paris, April 1995}
\fnm{1}{Also at the Laboratory of Theoretical Physics,
          JINR, Dubna, Russia}
\fnm{6}{Laboratoire associ\'e au Centre National de la Recherche
Scientifique (URA D0063)}
\section{Introduction}

The aim of this talk is to give a short review on the recent
developments in understanding the small$-x$ (and, in general, Regge)
asymptotics in perturbative QCD.

The present experimental data indicate a power-like bevahiour of the
structure functions of deep inelastic lepton scattering at small values
of Bjorken $x$,
$$
F_2(x,Q^2) \sim x^{-\varepsilon}\,, \hspace*{5mm}
\mbox{for $10^{-4} < x < 10^{-3}$}
$$
with $\varepsilon \approx 0.3$.
At small $x$, the structure function measures the total cross section of
deep inelastic scattering $\gamma^*(Q)+ {\rm p} \to X$ in the extreme
limit of high center-of-mass energies $s=Q^2(1-x)/x \gg Q^2$, in which
the famous Regge model emerges \ci{Col}. The Regge model interprets the
increasing of the structure function at high energies, or equivalently
at small $x$, by introducing the notion of the Pomerons as Regge poles
of the moments $\int_0^1 dx\, x^\omega F_2(x,Q^2)$
in the complex $\omega-$plane. The contribution of the Pomerons
to the structure function is given by
\be
F_2(x,Q^2) = \sum_{\tiny \IP} x^{-(\alpha_{\tiny \IP}-1)}\,
\beta_p^{\tiny \IP}\, \beta_{\gamma^*(Q^2)}^{\tiny \IP}
\lab{R}
\ee
where summation is performed over ``quantum numbers'' of the Pomerons $\IP$.
Here, $\alpha_{\tiny \IP}$ is the Pomeron intercept and $\beta_p^{\tiny
\IP}$ and $\beta_{\gamma^*}^{\tiny \IP}$ are the so-called residue
factors corresponding to proton and photon, respectively. Although the Regge
model gives a successful phenomenological description of the
experimental data in terms of ``hard''  (perturbative) and ``soft''
(nonperturbative) Pomerons \ci{softP} it is still unclear whether the
model is consistent with QCD.

The first attempts to understand the status of the ``hard''
Pomerons within perturbative QCD were undertaken more than 20 years ago
and they have led to the discovery of the BFKL Pomeron \ci{bfkl}.
The BFKL Pomeron was found in the leading logarithmic approximation
(LLA), $\as \ll 1$ and $\as \ln x \sim 1$, and at arbitrary small $x$
it leads to unrestricted rise of the structure function, or equivalently
of the parton densities in proton, and violates the unitary Froissart
bound \ci{Col}
\be
F_2(x,Q^2) < {\rm const.} \times \ln^2 x \hspace*{5mm}
\mbox{for $x\to 0$}\,.
\lab{F}
\ee
This means that at $x\to 0$ the BFKL Pomeron alone is not sufficient to
describe the small$-x$ asymptotics of the structure function.
One has to identify the ``nonleading'' Pomerons whose contribution
to the structure function is suppressed in the LLA by powers of $\as$
with respect to that of the BFKL Pomeron but which become important
for smaller values of $x$. In the next Sections the recent progress
on this problem will be reviewed.
\section{QCD Pomeron -- 20 years later}
\subsection{Perturbative QCD Pomeron}
Perturbative expansion of the structure function of DIS at small $x$
and fixed $Q^2$ has the following general form \ci{Bar}
\ba
F_2(x,Q^2)=\sum_{m=0}^\infty
\left[(\as\ln x)^m f_{m,m}(Q^2)
\right. \qquad
\nonumber
\\
\left.
+\as(\as\ln x)^{m-1} f_{m,m-1}(Q^2)+...
+\as^m f_{m,0}(Q^2)
\right]\,.
\lab{1}
\ea
Here, $\as \ll 1$ and $\as\ln x$ is a large parameter at small $x$.
The coefficient functions $f_{m,m-k}(Q^2)$ depend
on the internal structure of the proton and on the photon virtuality
$Q^2$.
It is clear that the number of different terms in \re{1} rapidly
increases to higher orders in $\as$ and in order to find the structure
function at small $x$ one has to develop a ``good'' approximation to
$F_2(x,Q^2)$ which, first, correctly describes the small$-x$
asymptotics of the infinite series \re{1} and, second, preserves the
unitarity constraint \re{F}.

To satisfy the first condition, one can neglect in \re{1}
the terms containing $f_{m,m-1},$ $...$, $f_{m,0}$
as suppressed by powers of $\as$ with respect to the leading term
$f_{m,m}$. The resulting series defines $F_2(x,Q^2)$ in the LLA
and it was resummed by BFKL \ci{bfkl} to all orders in $\as$
\be
F_2^{\rm LLA}(x,Q^2) = \sum_{m=0}^\infty  (\as\ln x)^m f_{m,m}(Q^2)
\sim x^{-\frac{\as N_c}{\pi} 4\ln 2}
\lab{BFKL}
\ee
with $N_c$ the number of quark colors. As was stressed before, this
expression violates the unitary bound \re{F}. In order to preserve
the unitarity of the S-matrix of QCD and fulfill \re{F} we have
to take into account an {\it infinite\/} number of nonleading terms
in \re{1}. This means that with the unitarity condition taken into
account the series \re{1} does not have any ``natural'' small parameter
of expansion like $\as$. However, instead of searching for this parameter
one may start with the LLA result \re{BFKL} and try to identify the
nonleading terms in \re{1}, which should be added to \re{BFKL} in order
to restore the unitarity. At present, the following three
unitarization schemes are known:

\medskip

\noindent
-- effective reggeon field theory \ci{Gr,Leff,VV}

\medskip

\noindent
-- generalized leading logarithmic approximation \ci{CW,Bar,KP}

\medskip

\noindent
-- dipole model \ci{onium}.

\medskip

\noindent
Each of them was developed to resum special class of corrections which
are expected to be dominated at small $x$. Namely, at small $x$ the
parton densities in parton increase and one has to deal with the dense
system of interacting partons. In this situation the description of
the process in terms of ``bare'' quarks and gluons becomes inappropriate
and one should instead identify a new collective degrees of freedom in
terms of which the dynamics becomes simpler. In the first two
unitarization schemes, the reggeized gluons (or Reggeons) \ci{bfkl} play
the role of such degrees of freedom while in the last scheme one deals
with the color dipoles \ci{onium}. The difference between the first two
schemes is that in the generalized LLA the unitarity is preserved only
in the direct channels of DIS but not in subchannels corresponding to
the different groups of particles the final state.
\subsection{Generalized leading logarithmic approximation}
In what follows we will analyze the structure function at small $x$
in the generalized LLA. Once we identified the Reggeons as a new
collective degrees of freedom in QCD at small $x$, we may try to develop
a new diagram technique for calculation of the structure function
\ci{CW,Bar,KP}.
Namely, an infinite set of standard Feynman diagrams involving ``bare''
gluons can be replaced by a few Reggeon diagrams describing propagation of
Reggeized gluons and their interaction with each other. Each Reggeon
diagram appears as a result of resummation of an infinite number of
Feynman diagrams with ``bare'' gluons.

In the generalized LLA the Reggeon diagrams have the form shown on
fig.~1.
Using the optical theorem, the structure function is defined as
an imaginary part of their contribution. Solid lines represent Reggeons
propagating in the $t-$channel and dotted lines denote interaction of
reggeized gluons. Upper and lower blobs describe the coupling of
Reggeons to the virtual photon, $\gamma^*(Q^2)$, and proton states,
respectively.

\vspace*{7mm}

%\unitlength=1.00mm
\unitlength=0.5mm
\linethickness{0.4pt}
\begin{picture}(130.00,128.00)
\put(70.50,107.50){\oval(81.00,15.00)[]}
\put(70.50,20.00){\oval(81.00,20.00)[]}
\put(111.00,22.00){\line(1,0){19.00}}
\put(130.00,22.00){\line(-1,0){19.00}}
\put(111.00,18.00){\line(1,0){19.00}}
\put(11.00,22.00){\line(1,0){19.00}}
\put(30.00,18.00){\line(-1,0){19.00}}
\put(17.00,23.00){\line(5,-3){5.00}}
\put(22.00,20.00){\line(-5,-3){5.00}}
\put(120.00,17.00){\line(2,1){6.00}}
\put(126.00,20.00){\line(-2,1){6.00}}
\put(13.00,28.00){\makebox(0,0)[cc]{$P$}}
\put(128.00,28.00){\makebox(0,0)[cc]{$P$}}
%\linethickness{1.5mm}
\linethickness{0.75mm}
\put(40.00,30.00){\line(0,1){70.00}}
\put(55.00,100.00){\line(0,-1){70.00}}
\put(70.00,30.00){\line(0,1){70.00}}
\put(85.00,100.00){\line(0,-1){70.00}}
\put(102.00,30.00){\line(0,1){70.00}}
\linethickness{1.4pt}
\put(40.00,103.00){\makebox(0,0)[cc]{$1$}}
\put(55.00,103.00){\makebox(0,0)[cc]{$2$}}
\put(70.00,103.00){\makebox(0,0)[cc]{$3$}}
\put(102.00,103.00){\makebox(0,0)[cc]{$k$}}
\put(85.00,103.00){\makebox(0,0)[cc]{$\cdots$}}
\linethickness{0.7pt}
\bezier{12}(40.00,78.00)(47.00,78.00)(53.00,78.00)
\bezier{12}(57.00,78.00)(63.00,78.00)(68.00,78.00)
\bezier{11}(72.00,78.00)(78.00,78.00)(83.00,78.00)
\bezier{13}(87.00,78.00)(94.00,78.00)(102.00,78.00)
\bezier{16}(55.00,90.00)(63.00,90.00)(70.00,90.00)
\bezier{16}(55.00,66.00)(48.00,66.00)(40.00,66.00)
\bezier{16}(70.00,60.00)(78.00,60.00)(85.00,60.00)
\bezier{16}(55.00,44.00)(48.00,44.00)(40.00,44.00)
\bezier{16}(55.00,53.00)(63.00,53.00)(70.00,53.00)
\bezier{16}(85.00,88.00)(94.00,88.00)(102.00,88.00)
\bezier{16}(85.00,35.00)(94.00,35.00)(102.00,35.00)
\bezier{16}(85.00,50.00)(94.00,50.00)(102.00,50.00)
\put(111.50,111.50){\oval(5,5)[rb]}
\put(116.50,111.50){\oval(5,5)[lt]}
\put(116.50,116.50){\oval(5,5)[rb]}
\put(121.50,116.50){\oval(5,5)[lt]}
\put(121.50,121.50){\oval(5,5)[rb]}
\put(126.50,121.50){\oval(5,5)[lt]}
\put(118.00,128.00){\makebox(0,0)[cc]{$\gamma^*$}}
\put(24.00,128.00){\makebox(0,0)[cc]{$\gamma^*$}}
\put(29.50,111.50){\oval(5.00,5.00)[lb]}
\put(24.50,111.50){\oval(5.00,5.00)[rt]}
\put(24.50,116.50){\oval(5.00,5.00)[lb]}
\put(19.50,116.50){\oval(5.00,5.00)[rt]}
\put(19.50,121.50){\oval(5.00,5.00)[lb]}
\put(14.50,121.50){\oval(5.00,5.00)[rt]}
\end{picture}

\noindent
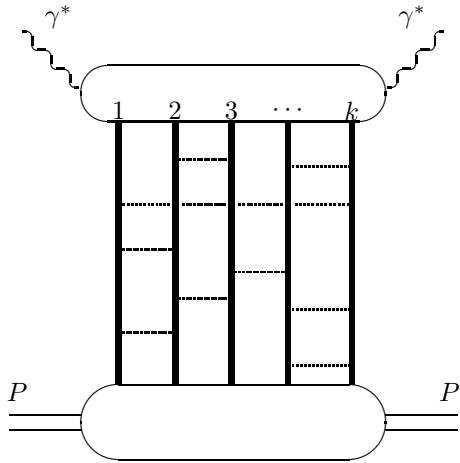
{\bf Figure 1:} Unitary Reggeon diagrams contributing to the structure
function of DIS at small $x$ in the generalized LLA in the multi-color
limit, $N_c\to \infty$. For finite $N_c$ one has to add similar
diagrams with pair-wise interaction between $k$ Reggeons.

\bigskip

The interaction between Reggeons is elastic
and pair--wise. This means that the number of Reggeons emitted by
proton is not changed and the diagrams of fig.~1 describe the propagation
in the $t-$channel of the conserved number $k=2,\ 3, ...$ of pair-wise
interacting Reggeons. As a result, the structure function is
given in the generalized LLA by a sum of $k-$Reggeon diagrams whose
contribution can be represented as
\be
F_2(x,Q^2) = \sum_{k=2}^\infty \as^{k-2}\ F^{(k)}(x,Q^2)\,,
\lab{3}
\ee
where $k$ is the conserved number of Reggeons in the $t-$channel
and the functions $F^{(k)}(x,Q^2)$ have the following form
\be
F^{(k)}(x,Q^2)=\sum_{m=k-2}^\infty (\as \ln x)^{m-k+2} f_{m,m-k+2}\,.
\lab{4}
\ee
Comparing \re{1} and \re{3} we notice that both expansions are similar and,
moreover, the nonleading corrections to the structure function can
be associated with the contribution of the $k-$Reggeon diagrams.
In the LLA, the functions
$F^{(k)}$ are of the same order for different $k$ and the contribution
of the $k-$Reggeon diagrams to \re{3} is suppressed by powers of $\as$
with respect to the leading term, $F^{(2)}$, which describes the propagation
of two Reggeized gluons. At smaller values of $x$, the growth
of the functions $F^{(k)}$ overwealms the suppression by
powers of $\as$ in \re{3}
and one has to take into account all functions $F^{(k)}$ in the
expansion \re{3}. This is a scenario of how nonunitarity of the LLA
result is restored in the generalized LLA.

The calculation of the $k-$Reggeon diagrams can be effectively
performed using the Bartels--Kwiecinski--Praszalowicz resummation
technique \ci{Bar,KP}. It turns out that for fixed number of Reggeons,
$k$, the functions $F^{(k)}$ satisfy Bethe--Salpeter like equation. Its
solutions have the following Pomeron like behaviour
$$
F^{(k)}(x,Q^2) \sim x^{-(\alpha_k-1)}\,, \hspace*{5mm} k=2,\,3,\,...
$$
which appears as a contribution of the compound states of
$k$ Reggeized gluons with vacuum quantum numbers, $\IP_k$, and
intercept $\alpha_k$.

Thus, the Pomerons naturally appear in the generalized
LLA as compound states of Reggeized gluons. The simplest example of
such states, the $k=2-$Reggeon state \ci{bfkl,Lip1}, defines the
structure function in the LLA, eq.\re{BFKL}, and it is identical to the
BFKL Pomeron with the intercept
\be
\alpha_2-1=\frac{\as N_c}{\pi} 4\ln2\,.
\lab{m1}
\ee
Finally, for the structure function of DIS we find the following
expression in the generalized LLA \ci{K}:
\be
F_2(x,Q^2)=\sum_{k=2}^\infty \as^{k-2} \sum_{\{q\}}
x^{-E_{k,\{q\}}}\, \beta_{\gamma^*(Q^2)}^{\{q\}}
\beta_p^{\{q\}}\,,
\lab{S}
\ee
where the first sum goes over all possible number $k$ of Reggeons
propagating in the $t-$channel and
the second summation is performed over all possible sets of the
quantum numbers ${\{q\}}$ parameterizing $k-$Reggeon compound states.
Here, the energy of the $k-$Reggeon compound state,
$E_{k,\{q\}}$, is defined as the
eigenvalue of the hamiltonian ${\cal H}_k$ describing a pair-wise
interaction of $k$ Reggeons \ci{Bar,KP}
\be
{\cal H}_k\ket{\chi_{k,\{q\}}}
= E_{k,\{q\}}\ket{\chi_{k,\{q\}}}\,.
\lab{Sc}
\ee
The eigenstate $\ket{\chi_{k,\{q\}}}$ is the wave function of the
Reggeon compound state and the residue factors
$$
\beta_{\gamma^*(Q^2)}^{\{q\}}=\langle \gamma^*(Q^2)
\ket{\chi_{k,\{q\}}}\,,\quad
\beta_p^{\{q\}}=\langle p \ket{\chi_{k,\{q\}}}
$$
measure its coupling to the photon and proton states.

The expression \re{S} is very similar to the predictions of the Regge
model \re{R} provided that we identify the intercept of the Pomerons
$\IP_k$ as the maximal energy of the $k$ Reggeon states
\be
\alpha_k-1={\rm max}_{\{q\}} E_{k,\{q\}}\,.
\lab{m2}
\ee

To find the structure function \re{S} one has to solve the Schr\"odinger
equation \re{Sc} for the system of $k$ Reggeons.
The hamiltonian ${\cal H}_k$ describes elastic
pair-wise interaction between $k$ Reggeons and it acts both on momenta
and color charges $t^a_j$ $(j=1,...,k)$ of the Reggeons
$$
{\cal H}_k=\frac{\as}{4\pi}\sum_{i,j=1;i>j}^k \, t^a_i t^a_j\ H_{ij}\,.
$$
Each Reggeon carries the same color charge as a ``bare'' gluon,
$(t^a)_{bc}=-if_{abc}$, while the
total color charge of $k$ Reggeons is preserved and it is equal to zero
for color singlet Pomeron state. Another remarkable property of Reggeon
hamiltonian is that the two-particle hamiltonian $H_{ij}$
changes only 2-dimensional transverse components of the Reggeon momenta and
it does not affect longtitudinal components \ci{bfkl}.

Thus, the equation \re{Sc} has a form of (2+1)--dimensional
Scr\"odinger equation for the system of $k$ particles with additional
color degrees of freedom. One should notice that this equation has been
solved in the special case of $k=2$ Reggeon states, or BFKL Pomeron
\ci{bfkl}, and
it remained unclear whether it can be solved for an arbitrary $k$.
Recently, a significant progress has been achieved \ci{Lip,FK} after it was
realized that in the multi--color limit, $N_c\to\infty$,
the Schr\"odinger equation \re{Sc} has an interesting interpretation in terms
of exactly solvable lattice models \ci{Q}.
\subsection{Multi-color limit}
The interaction between Reggeons can be simplified by taking the
multi-color limit \ci{Ven}, $\as N_c={\rm fixed}$ and $N_c\to\infty$,
and passing by
means of Fourier transformation from two-dimensional transverse momenta
of Reggeons to two-dimensional impact parameter space, $b_\bot=(b_1,b_2)$.
After these transformations, the Reggeon hamiltonian takes the following
form \ci{Lip2}
\be
{\cal H}_k = \frac{\as N_c}{4\pi}\lr{H_k + \bar H_k} + \CO(N_c^{-2})\,.
\lab{RH}
\ee
The hamiltonians $H_k$ and $\bar H_k$ act on
holomorphic and antiholomorphic coordinates of the Reggeons in the
impact parameter space, $z=b_1+ib_2$ and $\bar z=b_1-ib_2$, respectively,
and describe nearest--neighbour interaction between $k$ Reggeons
$$
H_k = \sum_{m=1}^k H(z_m,z_{m+1})\,,\qquad
\bar H_k = \sum_{m=1}^k H(\bar z_m,\bar z_{m+1})\,,
$$
where $z_m$ and $\bar z_m$ are holomorphic and antiholomorphic
coordinates of the $m-$th Reggeon and periodic boundary conditions
$z_{k+1}=z_1$ and $\bar z_{k+1}=\bar z_1$ are imposed.
The interaction hamilonian of two Reggeons with coordinates
$(z_1,\bar z_1)$ and $(z_2,\bar z_2)$ in the impact parameter space,
is given by the so--called BFKL kernel
\ci{Lip2},
\be
H(z_1,z_2)=-\psi(-J_{12})-\psi(1+J_{12})+2\psi(1)
\lab{psi}
\ee
where $\psi(x)=\frac{d}{dx}\ln\Gamma(x)$ and the operator $J_{12}$ is
defined as a solution of the equation
\be
J_{12} (1+J_{12}) = -(z_1-z_2)^2\partial_1\partial_2
\lab{J}
\ee
with $\partial_m=\partial/\partial z_m$. The expression
for $H(\bar z_1,\bar z_2)$ is similar to \re{psi}.

It follows from \re{RH} that in the multi--color limit
the 2--dimensional Reggeon
hamiltonian ${\cal H}_k$ describing the pair--wise interaction of
$k$ Reggeons turns out to be equivalent to the sum
of two 1--dimensional mutually commuting hamiltonians:
$[H_k, \bar H_k] = 0$.
This allows to reduce the original (2+1)--dimensional
problem \re{Sc} to the system of two (1+1)--dimensional Schrodinger
equations \ci{Lip2}
\ba
H_k \ket{\varphi_{k,\{q\}}} &=&
\varepsilon_{k,\{q\}} \ket{\varphi_{k,\{q\}}}\,,
\nonumber
\\
\bar H_k \ket{\bar\varphi_{k,\{q\}}} &=& \bar\varepsilon_{k,\{q\}}
\ket{\bar\varphi_{k,\{q\}}}\,,
\lab{H}
\ea
where the wave functions ${\varphi_{k,\{q\}}}$ and ${\bar\varphi_{k,\{q\}}}$
depend only on holomorphic and antiholomorphic coordinates of the Reggeons,
respectively. Then, in the multi--color limit the spectrum of the $k$
Reggeon states can be found as
\baa
&&
E_{k,\{q\}} = \frac{\as N_c}{4\pi}
\lr{\varepsilon_{k,\{q\}}+\bar\varepsilon_{k,\{q\}}}\,,
\\
&&
\ket{\chi_{k,\{q\}}} = \ket{\varphi_{k,\{q\}}}
\ket{\bar\varphi_{k,\{q\}}}
\,.
\eaa
We conclude that starting with the calculation of the structure function
of DIS at small $x$ in (3+1)--dimensional multi-color QCD in the
generalized LLA we came to the solution of the system of
(1+1)--dimensional Schr\"odinger equations \re{H}.
\section{Multi-color QCD at small $x$ as XXX Heisenberg magnet}
For fixed number of Reggeons, $k$, each of the Schr\"odinger equations
in \re{H} describes the system of $k$ one-dimensional particles on a line
interacting with their neighbours via hamiltonian \re{psi}.
It is of no surprise now that this quantum-mechanical system can
be exactly solved for $k=2$ particles leading to the BFKL Pomeron,
but it is not obvious that the same could be done for an arbitrary
number of Reggeons. The famous example of one-dimensional exactly
solvable multiparticle system is the XXX Heisenberg chain of $k$
interacting $s=1/2$ spins with the hamiltonian \ci{Q}
$$
H^{\rm XXX; s=1/2}_k = - \sum_{m=1}^k \vec S_m \vec S_{m+1}\,,
$$
where $\vec S_m$ are spin $s=1/2$ operators in the $m-$th site.
It turns out \ci{XXX} that this simple model admits nontrivial
generalizations to the XXX magnets for an arbitrary complex value of
the spin $s$. Moreover, in the special case $s=0$ the hamiltonian
of the XXX magnet turns out to be identical to the holomorphic
QCD hamiltonian defined in \re{psi}.

To explain this correspondence let us consider the one-dimensional
lattice with periodic boundary conditions and with the number of sites,
$k$, equal to the number of Reggeons. Each site is parameterized by
holomorphic coordinates $z_m$ $(m=1,...,k)$ and the spin $s=0$
operators are introduced in all sites as follows
$$
S^+_m= z_m^2 \partial_m\,,\quad S_m^-=-\partial_m\,,\quad
S_m^3=z_m \partial_m
$$
where $S^\pm=S^1\pm S^2$.
Then, the hamiltonian of exactly solvable XXX magnet of spin $s=0$
is defined as \ci{XXX,FK}
\ba
&&
H^{\rm XXX; s=0}_k=\sum_{m=1}^k H_{m,m+1}
\nonumber
\\
&&
H_{m,m+1}=-i\frac{d}{d\lambda} \ln R_{m,m+1}(\lambda)\bigg|_{\lambda=0}
\lab{X}
\ea
where the operator $R_{m,m+1}$, the so-called fundamental R-matrix,
satisfies the Yang-Baxter equation and it is given by \ci{R-m}
$$
R_{12}(\lambda)=
\frac{\Gamma^2(i\lambda+1)}{\Gamma(i\lambda-J_{12})\Gamma(i\lambda+J_{12}+1)}
$$
with $J_{12}$ introduced in \re{J}.
One can easily check that thus defined ``holomorphic'' two-particle
hamiltonian of the XXX Heisenberg magnet of spin $s=0$, \re{X}, coincides with
the holomorphic hamiltonian of Reggeon interaction \re{psi}. This
leads to the following identity between hamiltonians of two models
\be
H_k^{\rm Reggeon} \equiv H_k^{\rm XXX; s=0}
\lab{=}
\ee
which immediately implies that {\it the system of the Schr\"odinger equations
\re{H} describing the multi-color QCD Pomerons in the generalized LLA
is completely integrable\/} \ci{Lip,FK}.

Moreover, it follows from \re{=}, that the $k$ Reggeon compound states
share all their properties with the eigestates of the XXX Heisenberg
magnet for spin $s=0$. In particular, changing a sign of the Reggeon
hamiltonian \re{RH}
one could obtain \ci{K} that {\it the intercept of the $k-$Reggeon
states, \re{m2}, is a ground state energy of the XXX
magnet\/}. The latter can be found
%by applying the powerful methods of
%solving exactly solvable models, in particular
by using the Bethe Ansatz
technique \ci{BA1,BA2,BA3,BA4}. This program was initiated in \ci{FK,K}
where the generalized Bethe Ansatz was developed for diagonalization of
the Reggeon hamiltonian.
\section{Bethe Ansatz for QCD Pomerons}
The fact that the system of Schrodinger equations \re{H}
is completely integrable
implies that there exists a family of ``hidden'' holomorphic and
antiholomorphic conserved charges, $\{q\}$ and $\{\bar q\}$, which
commute
with the Reggeon hamiltonian \re{RH} and among themselves. Their explicit form
can
be found using the quantum inverse scattering method as \ci{Lip,FK}
\be
q_m=\sum_{k\ge i_1>i_2>\cdots >i_m\ge 1}
i^m z_{i_1i_2} z_{i_2i_3} \ldots z_{i_mi_1}
\partial_{i_1}\partial_{i_2} \ldots \partial_{i_m}
\lab{opqn}
\ee
with $z_{ij}\equiv z_i-z_j$
and the expression for $\bar q_m$ is similar. The appearance of these
operators is closely related to the invaiance of the Reggeon hamiltonian
\re{RH} under conformal $SL(2,\IC)$ transformations \ci{Lip1}
\be
z \to \frac{az+b}{cz+d}\,,\qquad
\bar z \to \frac{\bar a\bar z+ \bar b}{\bar c\bar z+\bar d}
\lab{ct}
\ee
where $ac-bd=\bar a\bar c-\bar b\bar d=1$. Indeed, we recognize $q_2$
and
$\bar q_2$ as quadratic Casimir operators of the $SL(2,\IC)$ group
while
the remaining conserved charges $\{q_m,\bar q_m\}$, $m=3,...,k$ can be
interpreted as higher Casimir operators. The Reggeon compound states
belong to the principal series representation of the $SL(2,\IC)$ group
and under the conformal transformations \re{ct} they are transformed as
quasiprimary fields with conformal weights $(h,\bar h)$ \ci{BPZ}.

The $k$ Reggeon states diagonalize the operators $\{q,\bar q\}$ and the
eigenvalues of the
conserved charges $q_2,$ $q_3$, $\ldots$, $q_k$ play a role of their
additional quantum numbers.
In particular, the eigenvalues of the quadratic Casimir operators
are related to the conformal weights of the $k$ Reggeon state as
$$
q_2=-h(h-1) \,,\qqquad
\bar q_2 =-\bar h (\bar h-1)\,,\qqquad
\bar q_2=q_2^*\,.
$$
where the possible values of $h$ and $\bar h=1-h^*$
can be parameterized by integer $n$ and
real $\nu$
\be
h=\frac{1+n}2 + i\nu\,,\qquad
n=\IZ\,,\quad \nu = \IR\,.
\lab{h}
\ee
As to remaining charges, $q_3$, ..., $q_k$, their possible values also
become quantized \ci{K}. The explicit form of the corresponding
quantization conditions is more complicated and it was obtained
in \ci{Qu}.

To find the explicit form of the eigenstates and eigenvalues of the
$k$ Reggeon compound states corresponding to a given set of quantum
numbers $\{q,\bar q\}$ we apply the generalized Bethe Ansatz
developed in \ci{FK,K}.
The Bethe Ansatz for Reggeon states in multi--color QCD is based on
the solution of the Baxter equation
\be
\Lambda(\lambda)
Q(\lambda)=(\lambda+i)^k Q(\lambda+i) + (\lambda-i)^k Q(\lambda-i)\,.
\lab{Bax}
\ee
Here, $Q(\lambda)$ is a real function of the spectral parameter
$\lambda$, $\Lambda(\lambda)$ is the eigenvalue of the so--called
auxiliary transfer matrix for the XXX Heisenberg magnet of spin $s=0$
\be
\Lambda(\lambda)=2\lambda^k+q_2 \lambda^{k-2} + \ldots + q_k
\lab{Lam}
\ee
and $k$ is the number of Reggeized gluon or, equivalently,
the number of sites of the one-dimensional spin chain.
For fixed $k$ it is convenient to introduce the function
\be
\widetilde Q(\lambda)=\lambda^k Q(\lambda)
\lab{tilde}
\ee
and rewrite the Baxter equation \re{Bax} as
\be
\Delta\, \widetilde Q(\lambda)
%(\lambda+i)+\widetilde Q(\lambda-i) - 2\widetilde Q(\lambda)
= \lr{
-\frac{h(h-1)}{\lambda^2}
+ \frac{q_3}{\lambda^3} + \ldots + \frac{q_k}{\lambda^k}}
\widetilde Q(\lambda)\,,
\lab{tildeBax}
\ee
where $\Delta$ is a second-order finite difference operator
$$
\Delta\, \widetilde Q(\lambda)=
\widetilde Q(\lambda+i)+\widetilde Q(\lambda-i) - 2\widetilde Q(\lambda)\,.
$$
Once we know the function $\widetilde Q(\lambda)$, the energy $E_k$ of
the $k-$Reggeon compound state can be evaluated using the relation
\ci{FK,K}
\be
E_k=\frac{\as N_c}{2\pi} \Re \varepsilon_k(h,q_3,\ldots,q_k)
\lab{En}
\ee
where the holomorphic energy $\varepsilon_k$ is defined as
\be
\varepsilon_k(h,q_3,\ldots,q_k)=i\frac{d}{d\lambda}
\log\frac{\widetilde Q(\lambda-i)}
         {\widetilde Q(\lambda+i)}\Bigg|_{\lambda=0}\,.
\lab{en}
\ee
The expression for the wave function of the $k$ Reggeon states in terms
of the function $\widetilde Q$ can be found in \ci{FK,K}.

The Baxter equation \re{tildeBax}
has the following properties \ci{K}. We notice that for
$q_k=0$ it is effectively reduced to a similar equation for the
states with $k-1$ Reggeized
gluons. The corresponding solution, $Q(\lambda)$, gives rise to the
degenerate
unnormalizable $k$ Reggeon states with the energy
\be
\varepsilon_k(h,q_3,...,q_{k-1},0)=
\varepsilon_{k-1}(h,q_3,...,q_{k-1})\,.
\lab{dege}
\ee
These states should be excluded from the spectrum of the $k$ Reggeon
hamiltonian and solving the Baxter equation for $k$ Reggeized gluons we
have to satisfy the condition
\be
q_k \neq 0\,.
\lab{neq}
\ee
As a function of the quantum numbers, the holomorphic energy obeys
the relations
\ba
\varepsilon_k(h,q_3,...,q_k)
&=&\varepsilon_k(1-h,q_3,...,q_k)
\nonumber
\\
&=&\varepsilon_k(h,-q_3,...,(-)^k q_k)\,,
\lab{q3-q3}
\ea
which follow from the symmetry of the Baxter equation \re{tildeBax}
under the
replacement $h\to 1-h$ or $\lambda\to-\lambda$ and $q_m\to (-)^m q_m$.
This relation means that the spectrum of the Reggeon hamiltonian
is degenerate with respect to quantum numbers $h$, $q_3$, $...$, $q_k$.
Then, assuming that the ground state of the XXX Heisenberg magnet of
spin $s=0$
is not degenerate we can identify the quantum numbers corresponding to
the maximal value of the Reggeon energy as \ci{K}
\be
h\bigg|_{\rm max} = \frac 12
\lab{1/2}
\ee
and for the states with only even number of Reggeons, $k=2m$,
$$
q_3\bigg|_{\rm max}=q_5\bigg|_{\rm max}=...=q_{2m-1}\bigg|_{\rm
max}=0\,.
$$
For the states with odd number of Reggeons the latter condition is not
consistent with \re{neq}.

We notice that the conformal weight $h$ enters as a parameter into the
Baxter equation \re{Bax} and, in general, one is interesting to find the
solutions of \re{Bax} only for its special values \re{h}.
In \ci{FK,K} the following way of solving the Baxter equation was
proposed. One first solves \re{Bax} for integer positive values of the
conformal weight $h$ and then analytically continues the result,
$Q(\lambda)$, to all possible values \re{h} including the most
physically interesting value \re{1/2}.
The Baxter equation \re{Bax} has two linear indepenent solutions and in
order to select only one of them we have to impose the additional
condition on the function $Q(\lambda)$ for $h=\IZ_+$
\be
Q(\lambda)\stackrel{\lambda\to\infty}{\sim} \lambda^{h-k}\,.
\lab{as}
\ee
For integer positive conformal weight, $h\ge k$, the solution,
$Q(\lambda)$, of the Baxter equation \re{Bax} under the additional
condition
\re{as} is given by a polynomial of degree $h-k$ in the spectral
parameter $\lambda$, which can be expressed
in terms of its roots $\lambda_1$, $...$,
$\lambda_{h-k}$ as follows \ci{FK,K}
\be
Q(\lambda) = \prod_{i=1}^{h-k} (\lambda-\lambda_i)\,.
\lab{roots}
\ee
Substituting \re{roots} into \re{tildeBax} and putting $\lambda=\lambda_i$
we obtain that the roots satisfy the famous Bethe equation for the XXX spin
chain
\be
\lr{\frac{\lambda_m+i}{\lambda_m-i}}^k
=\prod_{j=1 \atop j\neq m}^{h-k}
\frac{\lambda_m-\lambda_j-i}{\lambda_m-\lambda_j+i}\,,
\qquad m=1,...,h\,.
\lab{Be}
\ee
Unfortunately, there is no a regular way of solving the Bethe
equation \re{Be} for an arbitrary number $k$ of Reggeons.
At the same time, for $k=2$ Reggeon states the explicit solution of the
Baxter equation \re{Bax} was found in \ci{FK,K}
\be
Q_{k=2}(\lambda)= i^h h(1-h)\
{_3F_2}\left({1+h,2-h,1-i\lambda \atop 2, 2}; 1
\right)\,.
\lab{3F2}
\ee
where ${_3F_2}$ is the generalized hypergeometric function.
This expression admits an interesting interpretation \ci{K,Qu}
in terms of classical orthogonal polynomials and conformal field
theories. Substituting the solution \re{3F2} into \re{en} we obtain
the holomorphic energy of the $k=2$ Reggeon states as \ci{K}
\be
\varepsilon_2(h)=-4\left[\psi(h)-\psi(1)\right]
\lab{en-n=2}
\ee
where $\psi-$function was defined in \re{psi}. We substitute
$\varepsilon_2(h)$ into \re{En} and analytically continue the
result from integer $h$ to all possible complex values \re{h}
$$
E_2(h)=-2\frac{\as N_c}{\pi}
\Re\left[\psi\lr{\frac{1+|n|}2+i\nu}-\psi(1)\right]\,.
$$
This relation coincides with the well--known expression \ci{bfkl}
for the energy of
the $k=2$ Reggeon compound state, the BFKL Pomeron. The maximum
value of the energy
$$
E_2^{\rm max}=\frac{\as N_c}{\pi}4\ln 2
$$
is achieved at $h=1/2$ and it is in agreement with \re{m1}, \re{m2}
and \re{1/2}.

It is of most interest to find the solution of the Baxter equation
\re{Bax} for
higher $k\ge 3$ Reggeon states. Different approaches have been proposed
\ci{K,MW} but explicit expression similar to that \re{3F2} for the BFKL
Pomeron was not found yet.

It is not difficult however to solve the
Baxter equation \re{Bax} numerically for $k=3,\, 4\, ...$ Reggeon states
and for lowest values of the conformal weight $h$ and then find the
quantized charges $\{q_m\}$ and the energy $\varepsilon_k$.
The results of numerical solution of the Baxter equation for $k=3$ and
$k=4$ presented in \ci{K} indicate the remarkable regularity in the
distribution of the quantized values of charges $\{q_m\}$ and energy
$\varepsilon_k$ and they strongly suggest that the analytical solution
of the Baxter equation for high Reggeon states should exist. Recently,
such kind of solution was found in \ci{Qu} for an arbitrary number
of Reggeons, $k$, in the special limit of large values of the conformal
weight of the Reggeon states, $h\gg 1$.

The method developed in \ci{Qu} is based on the observation that after
rescaling of the spectral parameter $\lambda\to \lambda h$
in the Baxter equation \re{tildeBax}, the operator $\Delta$ can be
replaced in the naive limit $h\to \infty$ by a second-order derivative.
Then, the Baxter equation \re{tildeBax} takes a form of the
one-dimensional Schr\"odinger equation for a particle in external
potential, in which the inverse conformal weight $1/h$ of the Reggeon
states plays a role of the Planck constant. This fact allows us to
apply the well-known quasiclassical expansion and obtain the solution of
the Baxter equation as well as quantized values of the charges and
energy of $k$ Reggeon states
in the form of asymptotic series in $1/h$. For large integer
$h$ the obtained analytical expressions completely agree with the
results of numerical solutions while for small $h$ the asymptotic
expansions for the energy of $k$ Reggeon states becomes divergent
and it should be replaced by the asymptotic approximation. As first
nontrivial application of these results, the intercept of the $k=3$
Reggeon state, the so-called perturbative Odderon \ci{odd1}, was
obtained as \ci{Qu}
\be
\alpha_3-1 \le \frac{\as N_c}{\pi} 2.41\,.
\lab{Odd}
\ee
This relation estimates the Odderon intercept from above and it implies,
in particular, that it is smaller than the intercept \re{m1} of the
BFKL Pomeron. The expression \re{Odd} is also in agreement with the
lower bound for the Odderon intercept proposed in \ci{odd}.
\section{Summary}
It is still remains a challenge for QCD to understand the mechanism
responsible for the rise of the structure function of DIS at small
$x$. In this regime one has to deal with the system of strongly
correlated partons in proton, for which ``good'' standard methods like
operator product expansion are not applicable.

It is widely believed
that in the Regge limit and, in particular in the small$-x$ limit,
QCD should be replaced by a two-dimensional effective theory
(dual models, QCD string etc.), in which Reggeons play a role of a new
collective degrees of freedom. This theory should inherit all symmetries
of QCD and one may try to identify them by studying the small $x$
asymptotics of the structure function in DIS.

Indeed, analysing the small $x$ limit of the structure function of DIS in the
generalized LLA we found that multi-color QCD turns out to be equivalent
to the exactly solvable XXX Heisenberg magnet for noncompact spin $s=0$.
This relation still looks mysterious and one should understand better
its origin starting from QCD lagrangian. From practical point of view
it allows us to apply the powerful methods of exactly solvable models
for calculation the spectrum of Pomerons in perturbative QCD. The
work in this direction is at the very beginning and one should expect
a lot of surprises to come soon.

\vspace{1cm}
\Bibliography{100}

\bi{Col}  S.C. Frautschi, {\it Regge poles and S-matrix theory\/},
          New York, W.A. Benjamin, 1963;
\\        V. de Alfaro and T. Regge, {\it Potential scattering},
          Amsterdam, North-Holland, 1965;
\\        P.D.B. Collins, {\it An introduction to Regge theory
          and high energy physics\/}, Cambridge University Press, 1977.
\bi{softP}P.V. Landshoff, DAMTP preprint, Oct. 94,
          [hep-ph/9410250].
\bi{bfkl} E.A. Kuraev,  L.N. Lipatov and V.S. Fadin,
          Phys. Lett. B60 (1975) 50;
          Sov. Phys. JETP 44 (1976) 443; 45 (1977) 199;
\\        Ya.Ya. Balitsky and L.N. Lipatov, Sov. J. Nucl. Phys. 28 (1978) 822.
\bi{Bar}  J. Bartels, Nucl. Phys. B175 (1980) 365.
\bi{Gr}   V.N. Gribov, Sov. Phys. JETP 26 (1968) 414.
\bi{Leff} R. Kirschner, L.N. Lipatov and L. Szymanowski,
          Nucl. Phys. B425 (1994) 579.
\bi{VV}   E. Verlinde and H. Verlinde, Princeton Univ. preprint,
          PUPT--1319, Sept. 1993.
\bi{CW}   H. Cheng, J. Dickinson, C.Y. Lo and K. Olaussen,
          Phys. Rev. D23 (1981) 534;
\\        H. Cheng and T.T. Wu, {\it Expanding Protons: Scattering at
          High Energies\/}, MIT Press, Cambridge, Massachusetts, 1987.
\bi{KP}   J. Kwiecinski and M. Praszalowicz, Phys. Lett. B94 (180) 413.
\bi{onium}A.H. Mueller and B. Patel, Nucl. Phys. B425 (1994) 471;
\\        A.H. Mueller, Nucl. Phys. B415 (1994) 373; B437 (1995) 107;
\\        N.N. Nikolaev, B.G.Zakharov and V.R. Zoler, Phys. Lett. B328
          (1994) 486.
\bi{Lip1} L.N. Lipatov, {\it Pomeron in quantum chromodynamic\/},
          in ``Perturbative QCD'', pp.411--489, ed. A.H. Mueller,
          World Scientific, Singapore, 1989.
\bi{K}    G.P. Korchemsky, Nucl. Phys. B443(1995) 255.
\bi{Lip}  L.N. Lipatov, JETP Lett. 59 (1994) 596.
\bi{FK}   L.D. Faddeev and G.P. Korchemsky,
          Stony Brook preprint ITP--SB--94--14, Apr. 1994, [hep-ph/9404173];
          Phys. Lett. B 342 (1995) 311.
\bi{Q}    R.J. Baxter, {\it Exactly Solved Models in Statistical
          Mechanics\/}, Academic Press, London, 1982.
\bi{Ven}  G. Veneziano, Nucl. Phys. B74 (1974) 365;
          Phys. Lett. 52B (1974) 220;
\\        A. Schwimmer and G. Veneziano, Nucl. Phys. B81 (1974) 445;
\\        M. Cafaloni, G. Marchesini and G. Veneziano, Nucl. Phys. B98 (1975)
          472; 493.
\bi{Lip2} L.N. Lipatov, Phys. Lett. B251 (1990) 284;  B309 (1993) 394.
\bi{XXX}  V.O. Tarasov, L.A. Takhtajan and L.D. Faddeev, Theor. Math.
          Phys. 57 (1983) 163.
\bi{R-m}  P.P. Kulish, N.Yu. Reshetikhin and E.K. Sklyanin, Lett. Math.
          Phys. 5 (1981) 393.
\bi{BA1}  E.K. Sklyanin, L.A. Takhtajan and L.D.Faddeev,
          Theor. Math. Phys. 40 (1980) 688.
\bi{BA2}  L.A. Takhtajan and L.D. Fddeev, Russ. Math. Survey 34 (1979) 11.
\bi{BA3}  L.D. Faddeev,
          Stony Brook preprint, ITP-SB-94-11, Mar 1994;hep-th/9404013;
          in Nankai Lectures on Mathematical Physics,
          Integrable Systems, ed. by X.-C.Song, pp.23-70,
          Singapore: World Scientific, 1990.
\bi{BA4}  V.E. Korepin, N.M.Bogoliubov and A.G. Izergin, {\it Quantum
          inerse scattering method and correlation functions\/},
          Cambridge Univ. Press, 1993.
\bi{BPZ}  A.A. Belavin, A.M. Polyakov and A.B. Zamolodchikov,
          Nucl. Phys. B241 (1984) 333.
\bi{Qu}  G.P. Korchemsky, Stony Brook preprint ITP--SB--95--25, July
         1995, [hep-th/9508025].
\bi{MW}   Z. Maassarani and S. Wallon, Saclay preprint,
          SACLAY-SPHT-95-081,
          Jun 1995, [hep-th/9507056].
\bi{odd1} P. Gauron and B. Nicolescu, Phys. Lett. B260 (1991) 40;
\\        P. Gauron, L. Lukaszuk and B. Nicolescu, Phys. Lett. B294
          (1992) 298;
\\        B. Nicolescu, Nucl. Phys. (Proc.Suppl.) 25B (1992) 142.
\bi{odd}  P. Gauron, L.N. Lipatov and B. Nicolescu,
          Z. Phys. C63 (1994) 253; Phys. Lett. B304 (1993) 334.
\end{thebibliography}
\end{document}